\documentclass[preprintnumbers,amsmath,amssymb,aps,prl]{revtex4}

\usepackage{graphicx}
\usepackage{dcolumn}
\usepackage{bm}
\usepackage{verbatim}
\begin{document}
\title{Room temperature spin filtering in epitaxial cobalt-ferrite tunnel barriers}

\author{A. V. Ramos}
\affiliation{DSM/DRECAM/SPCSI, CEA-Saclay, 91191 Gif-Sur-Yvette
Cedex, France}
\author{M.-J. Guittet} \affiliation{DSM/DRECAM/SPCSI,
CEA-Saclay, 91191 Gif-Sur-Yvette Cedex, France}
\author{J.-B. Moussy} \affiliation{DSM/DRECAM/SPCSI,
CEA-Saclay, 91191 Gif-Sur-Yvette Cedex, France}

\author{R. Mattana}
\affiliation{Unit\'{e} Mixte de Physique CNRS/Thales, Route
d\'{e}partementale 128, 91767 Palaiseau Cedex, and Universit\'{e}
Paris-sud, 91405 Orsay, France}
\author{C. Deranlot}
\affiliation{Unit\'{e} Mixte de Physique CNRS/Thales, Route
d\'{e}partementale 128, 91767 Palaiseau Cedex, and Universit\'{e}
Paris-sud, 91405 Orsay, France}
\author{F. Petroff}
\affiliation{Unit\'{e} Mixte de Physique CNRS/Thales, Route
d\'{e}partementale 128, 91767 Palaiseau Cedex, and Universit\'{e}
Paris-sud, 91405 Orsay, France}

\author{C. Gatel} \affiliation{CEMES/CNRS, 31055 Toulouse, France}

\date{\today}

\begin{abstract}
We report direct experimental evidence of room temperature spin
filtering in magnetic tunnel junctions (MTJs) containing
CoFe$_2$O$_4$ tunnel barriers via tunneling magnetoresistance (TMR)
measurements.
Pt(111)/CoFe$_2$O$_4$(111)/$\gamma$-Al$_2$O$_3$(111)/Co(0001) fully
epitaxial MTJs were grown in order to obtain a high quality system,
capable of functioning at room temperature. Spin polarized transport
measurements reveal significant TMR values of -18\% at 2 K and -3\%
at 290 K. In addition, the TMR ratio follows a unique bias voltage
dependence that has been theoretically predicted to be the signature
of spin filtering in MTJs containing magnetic barriers.
CoFe$_2$O$_4$ tunnel barriers therefore provide a model system to
investigate spin filtering in a wide range of temperatures.

\end{abstract}

\pacs{75.47.Pq, 85.75.-d, 72.25-b, 73.40.Rw}

\maketitle

\newpage

The generation of highly spin-polarized electron currents is one of
the dominant focusses in the field of spintronics. For this purpose,
spin filtering is one very interesting phenomenon, both from a
fundamental and from a technological stand point, that involves the
spin-selective transport of electrons across a magnetic tunnel
barrier. Successful spin filtering at room temperature could
potentially impact future generations of spin-based device
technologies \cite{In-SPIN-Fiederling-nature402,
BM-Moodera-j.phys.cond.mat19} not only because spin filters may
function with 100\% efficiency \cite{BM-Moodera-prl70}, but they can
be combined with any non-magnetic metallic electrode, thus providing
a versatile alternative to half-metals or MgO-based classic tunnel
junctions.

The spin filter effect originates from the exchange splitting of the
energy levels in the conduction band of a magnetic insulator. As a
consequence, the tunnel barrier heights for spin-up and spin-down
electrons ($\Phi_{\uparrow(\downarrow)}$) are not the same, leading
to a higher probability of tunneling for one of the two spin
orientations : $J_{\uparrow(\downarrow)} \propto
exp(-\Phi_{\uparrow(\downarrow)}^{1/2}t)$, where $t$ is the barrier
thickness. The spin filter effect was first demonstrated in EuS
using a superconducting electrode as a spin analyzer (i.e.
Merservey-Tedrow technique) \cite{SP-Meservey-phys.repts238}, and
has since been observed in EuSe \cite{BM-Moodera-prl70} and EuO
\cite{BM-Santos-prb69} by this method. Because the Merservey-Tedrow
technique is limited to low temperatures, tunneling
magnetoresistance (TMR) measurements in magnetic tunnel junctions
(MTJs) \cite{BM-LeClair-apl} have been used to show the spin filter
capability of higher temperature spin filters such as BiMnO$_3$
\cite{BM-Gajek-prb} and NiFe$_2$O$_4$ \cite{BM-Luders-apl88}.
However, no TMR effects are currently reported from any spin filter
at room temperature.

CoFe$_2$O$_4$ is a very promising candidate for room temperature
spin filter applications thanks to its high Curie temperature ($T_C$
= 793 K) and good insulating properties. Electronic band structure
calculations from first principles methods predict CoFe$_2$O$_4$ to
have a band gap of 0.8 eV, and an exchange splitting of 1.28 eV
between the minority (low energy) and majority (high energy) levels
in the conduction band \cite{CFO_CALC-Szotek-prb74} (see
Fig.\ref{fig:IV-TMR(V)}-b), thus confirming its potential to be a
very efficient spin filter, even at room temperature. Recently, a
tunneling spectroscopy study of
CoFe$_2$O$_4$/MgAl$_2$O$_4$/Fe$_3$O$_4$ double barrier tunnel
junctions revealed optimistic results for the spin-filter efficiency
of CoFe$_2$O$_4$ \cite{CFO_SF-Chapline-prb74}. However, the
polarization ($P$) and TMR values obtained in this work were
indirectly extracted from a complex model developed to fit
experimental current-voltage curves rather than from direct
Merservey-Tedrow or TMR measurements.

In order to accurately demonstrate the spin filtering capabilities
of CoFe$_2$O$_4$ up to room temperature, we have prepared
CoFe$_2$O$_4$(111)/$\gamma$-Al$_2$O$_3$(111)/Co(0001) fully
epitaxial tunnel junctions by oxygen plasma-assisted molecular beam
epitaxy (MBE) on Pt(111) underlayers. The details of the sample
growth process are published elsewhere \cite{CFO_SF-Ramos-prb75}. In
this system, the spinel $\gamma$-Al$_2$O$_3$ serves to decouple the
CoFe$_2$O$_4$ and Co magnetic layers. Before any spin-polarized
transport measurements were performed on the full MTJ system, our
CoFe$_2$O$_4$(111)/$\gamma$-Al$_2$O$_3$(111) tunnel barriers were
carefully characterized by a wide range of techniques in order to
optimize their structural and chemical properties
\cite{CFO_SF-Ramos-prb75}. Fig. \ref{fig:TEM} shows a high
resolution transmission electron microscopy (HRTEM) study
demonstrating the high crystalline quality of our CoFe$_2$O$_4$ (5
nm)/$\gamma$-Al$_2$O$_3$ (1.5 nm)/Co (10 nm) multilayers. In
particular, we observe near perfect epitaxy in the single
crystalline CoFe$_2$O$_4$(111)/$\gamma$-Al$_2$O$_3$(111) tunnel
barrier, which is a consequence of the optimized growth conditions
and the spinel structure of both constituents.

The magnetic properties of a CoFe$_2$O$_4$/$\gamma$-Al$_2$O$_3$
tunnel barrier were measured at room temperature and at 4 K,
yielding coercivities ($H_c$) of 220 Oe and 500 Oe respectively (not
shown). We also observed a rather weak remanent magnetization around
40\% and lack of saturation and magnetic reversibility for fields as
high as 5 T at 4 K. These properties are common for spinel ferrite
thin films and have been attributed to the presence of antiphase
boundaries \cite{CM-MAG-Margulies-prl}. We note the magnetic
properties of the full Pt/CoFe$_2$O$_4$/$\gamma$-Al$_2$O$_3$/Co MTJs
were also characterized in order to verify the magnetic decoupling
of the CoFe$_2$O$_4$ and Co layers, necessary for TMR measurements
\cite{CFO_SF-Ramos-prb75}.

Spin filter tunnel junctions were patterned by advanced optical
lithography. Spin-polarized transport measurements were carried out
in the two probe configuration, as the room temperature junction
resistance was 3 orders of magnitude higher than that of the Pt
cross strip and contacts. Fig.\ref{fig:TMR}-a,b clearly demonstrates
the TMR effect in a Pt/CoFe$_2$O$_4$/$\gamma$-Al$_2$O$_3$/Co tunnel
junction both at 2 K and 290 K. This result is direct experimental
evidence of the spin filter effect in CoFe$_2$O$_4$ at low
temperature and \emph{at room temperature}. At 2 K we calculate a
TMR value of $-18\%$ using the relation TMR = $(R_{AP}-R_{P})/R_{P}$
where $R_{AP}$ and $R_{P}$ are the resistance values in the
antiparallel and parallel magnetic configurations. At room
temperature, TMR = -3\%.  The abrupt drop in the TMR curve at
$\pm$200 Oe corresponds to the switching of the Co electrode, while
the gradual increase back to $\pm$6 T agrees with the progressive
switching and lack of saturation in CoFe$_2$O$_4$ seen in the
magnetization measurements. The negative sign of the TMR indicates
that the CoFe$_2$O$_4$ spin filter and Co electrode are oppositely
polarized which is most consistent with the negative $P$ predicted
for CoFe$_2$O$_4$ in band structure calculations
\cite{CFO_CALC-Szotek-prb74}, and the positive $P$ measured for Co
by the Meservey-Tedrow technique. Taking P$_{Co}$=40\%
\cite{SP-Moodera-jmmm200}, we may approximate P$_{CoFe_2O_4}$ from
Julli\`{e}re's formula: $TMR = 2P_1P_2/(1-P_1P_2)$ where
P$_1$=P$_{Co}$ and P$_2$=P$_{CoFe_2O_4}$
\cite{TMR-Julliere-phys.lett.A}. This gives P$_{CoFe_2O_4}$=-25\% at
2K and P$_{CoFe_2O_4}$=-4\% at room temperature. Due to the low
remanence in our CoFe$_2$O$_4$ films, one could expect these values
to increase significantly with the future improvement of their
magnetic properties, reaching 50\% or higher. The considerable
decrease of P$_{CoFe_2O_4}$ at high temperature could be explained
by the thermal excitation of the spin up electrons into their
corresponding majority spin conduction band, if the conduction band
splitting were small with respect to $k_BT$.

The $I-V$ characteristics representative of the
Pt/CoFe$_2$O$_4$/$\gamma$-Al$_2$O$_3$/Co system are shown in
Fig.\ref{fig:TMR}-c. Analyzing the second derivative of these
tunneling spectroscopy measurements, we obtain a good estimation of
$\Phi$ from the bias voltage at which these deviate from linearity.
Fig.\ref{fig:TMR}-d clearly shows that $d^2I/dV^2$ is linear for -60
mV $< V<$ 60 mV, indicating direct tunneling in this regime. We also
obtain the same value of $\Phi$ from the $(dI/dV)/(I/V)$
characteristics which show a peak at 60 mV corresponding to the
onset of the conduction band. We note that this $\Phi$ value does
not account for the voltage drop across the $\gamma$-Al$_2$O$_3$
barrier, which should in fact lower it further. In either case, the
relatively small tunnel barrier height is quite consistent with the
electronic band structure calculations schematized in
Fig.\ref{fig:IV-TMR(V)}-b, which predict CoFe$_2$O$_4$ to have a
small electronic band gap and an intrinsic Fermi level that is close
to the first level of the conduction band.

In addition, the $I-V$ curves taken in the antiparallel ($\pm$0.08
T) and parallel ($\mp$6 T) states may be used to extract the TMR
bias dependence by using the definition TMR =
$(I_{P}-I_{AP})/I_{AP}$. The result, shown in
Fig.\ref{fig:IV-TMR(V)}-a, is a steady increase in absolute value of
the TMR with increasing $V$ up to a certain value, followed by a
slight decrease for higher biases. This exact behavior was
theoretically predicted by A. Saffarzadeh to be the signature of
spin filtering in MTJs containing a magnetic barrier
\cite{CALC-Saffarzadeh-jmmm269}, and has only recently been observed
experimentally with EuS at low temperature \cite{BM-Nagahama-prl99}.
However, it has never been verified in higher $T_C$ magnetic oxide
tunnel barriers \cite{BM-Gajek-prb, BM-Luders-apl88}. The fact that
the TMR increases with increasing $V$ both at low temperature and at
room temperature proves that the TMR we observe truly results from
spin filtering across the CoFe$_2$O$_4$ barrier. The onset of spin
filtering, given by the region for which TMR increases with $V$, may
be identified from the low temperature curve in
Fig.\ref{fig:IV-TMR(V)}-a around $\pm$30 mV and persists until +130
mV (-100 mV) for positive (negative) bias.

Quantitative comparison of the spin filter regime in our TMR($V$)
curves with the calculations of Szotek \emph{et al.} yields an
exchange splitting (in the tens of meV in our junctions) that is
significantly lower than that predicted for the inverse spinel
structure (1.28 eV). This observation is consistent with the
temperature sensitivity of the TMR measurement discussed earlier.
The electronic band structure of the CoFe$_2$O$_4$ barrier is likely
influenced by the presence of structural and/or chemical defects,
many of which are difficult to account for in model systems for such
calculations. The presence of Co$^{3+}$, for example, is one defect
that has been predicted by Szotek \emph{et al.}
\cite{CFO_CALC-Szotek-prb74} to reduce the band gap as well as the
conduction band splitting while favoring an energetically stable
state. Furthermore, one can not ignore the possible influence of
oxygen vacancies or antiphase boundaries, as these are known to
influence the magnetic order in spinel ferrites
\cite{CM-MAG-Margulies-prl}, and thus likely the exchange splitting
as well. Further studies are underway to better quantify the effects
of both structural and chemical defects on the spin polarized
tunneling across CoFe$_2$O$_4$.

In summary, we have demonstrated room temperature spin filtering in
fully epitaxial
Pt(111)/CoFe$_2$O$_4$(111)/$\gamma$-Al$_2$O$_3$(111)/Co(0001) MTJs
where CoFe$_2$O$_4$ was the magnetic tunnel barrier. TMR values of
-18\% and -3\% were observed at 2 K and 290 K respectively.
Furthermore, the experimental TMR ratio increased with increasing
bias voltage, reproducing the theoretically predicted behavior for a
model spin filter system. The similarity between our experimental
TMR($V$) curves and those previously predicted in the literature not
only proves the spin filtering capability of CoFe$_2$O$_4$, but also
validates the theoretical and phenomenological models describing
spin-polarized tunneling across a magnetic insulator.

The authors are very grateful to M. Gautier-Soyer for helpful
discussions. This work is funded by CNANO \^{I}le de France under
the ``FILASPIN" project.


\newpage
\section{Figure Captions}

\begin{figure}[!h]
\caption{HRTEM image of a CoFe$_2$O$_4$ (5 nm)/$\gamma$-Al$_2$O$_3$
(1.5 nm)/Co (10 nm) trilayer deposited directly on a sapphire
substrate, and showing the exceptional quality of the fully
epitaxial system.} \label{fig:TEM}
\end{figure}

\begin{figure}[!h]
\caption{TMR as a function of applied magnetic field for a Pt(20
nm)/CoFe$_2$O$_4$(3 nm)/$\gamma$-Al$_2$O$_3$(1.5 nm)/Co(10 nm)
tunnel junction at 2 K (a) and at room temperature (b) with an
applied bias voltage of 200 mV. The junction area $A$ was 24
$\mu$m$^2$. A zoom of the Co switching at room temperature is shown
in the insert of (b). The $I-V$ characteristics, and $d^2I/dV^2$
fitted linearly for $-\Phi \leq V \leq \Phi$ are shown in (c) and
(d).} \label{fig:TMR}
\end{figure}

\begin{figure}[!h]
\caption{(a) TMR as a function of bias voltage for a Pt(20
nm)/CoFe$_2$O$_4$(3 nm)/$\gamma$-Al$_2$O$_3$(1.5 nm)/Co(10 nm)
tunnel junction at 2 K and 300 K. The open data points correspond to
TMR values obtained from $R(H)$ measurements. (b) Schematic
representation of the CoFe$_2$O$_4$ band structure based on first
principles studies \cite{CFO_CALC-Szotek-prb74}.}
\label{fig:IV-TMR(V)}
\end{figure}

\end{document}